# System Identification via Polynomial Transformation Method

Prof Pradip Sircar

Department of Electrical Engineering

Indian Institute of Technology Kanpur

Email address: sircar@iitk.ac.in

**Abstract**

We propose a method based on minimum-variance polynomial approximation to extract system poles from a data set of samples of the impulse response of a linear system. The method is capable of handling the problem under general conditions of sampling and noise characteristics. The superiority of the proposed method is demonstrated by statistical comparison of its performance with the performances of two exiting methods in the special case of uniform sampling.

**Introduction**

In the all-pole model for linear systems, the characterization of the impulse response by a sum of weighted complex exponential signals, and then estimating the complex frequency- and amplitude-parameters of the signal is equivalent to determining the poles and residues at the poles, respectively, of the system. This particular problem arises in the diversified fields of control, electromagnetics, geophysics, NMR spectroscopy, etc., and is, therefore, of eminent importance.

The parameter estimation of complex exponential signals, when the noise corrupted signal samples are provided at non-uniform spacing, has been attempted in [1], [2]. This problem has special significance because non-uniform sampling can be preferred over uniform sampling for various considerations like sampling efficiency, SNR enhancement on the sampled set of signal values, etc.

To deal with the non-uniform sampling case, the idea of applying the orthogonal polynomial approximation together with the minimum error-variance criterion to find a closed-form expression for the approximating signal has been introduced in [1], [2]. The closed-form expression can be utilized to reconstruct the signal values at uniform spacing.

In an extension of that work, it is presented in [3], [4] that reconstruction of signal values at uniform spacing is through a linear causal transformation based on minimum-variance polynomial approximation. Therefore, the statistical properties of the approximating error sequence in the reconstructed signal values can be obtained from the known statistics of the original noise process. We assume time-invariance of the transformation here.

The maximum likelihood estimator combined with a rank-reduced SVD algorithm can then be employed to obtain estimation with high accuracy in the general case of non-uniform sampling. The most desirable feature of the proposed method is that it provides estimation in the maximum likelihood sense even when the corrupting noise process is not white or Gaussian.

It is the purpose of this article to provide a statistical comparison of the performance of the proposed method with the performances of two existing SVD-based techniques. Our study shows the proposed method provides better accuracy of estimation over the existing methods even for the case where the noise process is white and Gaussian.

**Non-Uniform Sampling and Estimation of Parameters of Complex Exponential Signals**

The $s_i$-parameters of the signal modeled as

$$g(t) = \sum_{i=1}^{M} \exp(s_i t), \quad t \geq 0; \quad s_i = -\alpha_i + j2\pi f_i \quad (1)$$

are to be estimated by utilizing the noise corrupted signal samples $\{x(t_k) = g(t_k) + w(t_k); t_k = t_1, t_2, \cdots, t_K\}$ at nonuniform spacing, where $\{w(t_k)\}$ is the sequence sampled from a zero-mean noise process $w(t)$ whose normalized autocorrelation functions are known.

By employing the orthogonal polynomial approximation and minimum error-variance criterion, the reconstructed signal vector, $Y_L = [y(0), y(T), \cdots, y((L-1)T)]^T$ is computed as [4],

$$Y_L = \mathbf{PQ}^{-1}\mathbf{P}_1^T X_K \quad (2)$$

where $X_K = [x(t_1), x(t_2), \cdots, x(t_K)]^T$,

$\mathbf{P}_1 = [p_{j-1}(t_i); i = 1, \cdots, K; j = 1, \cdots, N]$,

$\mathbf{Q} = \text{diag}\left[ \sum_{k=1}^{K} p_0^2(t_k), \cdots, \sum_{k=1}^{K} p_{N-1}^2(t_k) \right]$,

$\mathbf{P} = [p_{j-1}((i-1)T); i = 1, \cdots, L; j = 1, \cdots, N]$,

and the polynomials $p_j(t)$ are given by

$$p_j = (t - a_j)p_{j-1} - b_j p_{j-2}, \quad j \geq 1 \quad (3)$$

with $p_0 = 1$, $p_{-1} = 0$,

and $a_j = (1/\Phi_{j-1})\sum_{k=1}^{K} t_k [p_{j-1}(t_k)]^2$, $b_j = \Phi_j / \Phi_{j-1}$, $\Phi_j = \sum_{k=1}^{K} [p_j(t_k)]^2$.

The approximation order $N$ is determined so that the error variance

$$\sigma_N^2 = \sum_{k=1}^{K} [x(t_k) - y(t_k)]^2 \Big/ (K - N) \quad (4)$$

is minimum. The sampling interval $T$ is utilized to convert the $z$-plane poles into the $s$-plane poles as,

$$s_i = \left(0.5\ln|z_i|^2 + j\arg(z_i)\right)/T \quad (5)$$

In order to apply the statistics of the approximating error sequence $\{e(iT) = y(iT) - g(iT)\}$ into the estimation procedure, it is observed that the transformation $(\mathbf{PQ}^{-1}\mathbf{P}_1^T)$ in (2) is linear, causal, and its impulse response sequence $\{h(iT)\}$ is obtained in the first column of the transformation matrix. We assume that the system is time-invariant.

It is then easy to visualize that the noise sequence $\{w(t_k)\}$ is converted into the error sequence $\{e(iT)\}$ by a linear system $\{h(iT)\}$. As a consequence, since the noise sequence is zero-mean, the error sequence is also zero-mean; furthermore, the autocorrelation functions are related by

$$r_{ee}[k] = h[k] ** h^*[-k] ** r_{ww}[k] \quad (6)$$

where $[k] = (kT)$, $^*$ stands for complex conjugation, and $**$ denotes discrete convolution.

The probability density function of $\{e(iT)\}$ will be jointly Gaussian when the noise process $w(t)$ is Gaussian. In fact, by invoking the central limit theorem, it can be shown that the error sequence will be close to Gaussian when $w(t)$ is white, but may not be Gaussian [4].

By employing the maximum likelihood estimator (MLE) in the linear prediction model well known for the form of signal expressed in (1), the prediction coefficients $\{-a_i\}$ are obtained by solving the following equation [4], [5],

$$\left(\mathbf{G}^H \mathbf{R}_{ee}^{-1} \mathbf{G}\right) A = \mathbf{G}^H \mathbf{R}_{ee}^{-1} Y \quad (7)$$

where $Y = \left[y[J], y[J+1], \cdots, y[L-1]\right]^T$,

$A = \left[-a_1, -a_2, \cdots, -a_J\right]^T$,

$\mathbf{G} = \left[g[J+i-j-1]; i=1,\cdots,L-J; j=1,\cdots,J\right]$,

$\mathbf{R}_{ee} = \left[r_{ee}[i-j]; i=1,\cdots,L-J; j=1,\cdots,L-J\right]$,

and, $J$ is the extended model order, $J \leq L - J$, and preferably, $J$ is much larger than $M$.

To compute $A$ from (7), the pseudoinverse of $\left(\mathbf{G}^H \mathbf{R}_{ee}^{-1} \mathbf{G}\right)$ is needed. The characteristic equation is then formed as

$$z^J + a_1 z^{J-1} + \cdots + a_J = 0 \quad (8)$$

which is solved to find $M$ signal and $(J-M)$ noise $z_i$-poles.

In practice, the matrix $\mathbf{G}$ cannot be formed because of unavailability of $\{g[i]\}$. Hence, (7) is formed by $\mathbf{Y}$ replacing $\mathbf{G}$,

$$\left(\mathbf{Y}^H \mathbf{R}_{ee}^{-1} \mathbf{Y}\right) A = \mathbf{Y}^H \mathbf{R}_{ee}^{-1} Y \quad (9)$$

where $\mathbf{Y} = \left[ y[J+i-j-1]; i=1,\cdots,L-J; j=1,\cdots,J \right]$.

In this case, the rank-$M$ pseudoinverse of $\left(\mathbf{Y}^H \mathbf{R}_{ee}^{-1} \mathbf{Y}\right)$ is employed to compute $A$, and the SVD technique is conveniently utilized for the purpose [6]. To get accurate results at very low SNR (say, 5 dB), corrections of the principal singular values are necessary [4].

**Uniform Sampling and Other SVD-based Methods**

Uniform sampling is the special case of nonuniform sampling. As such, the performance of the developed method which is based on minimum-variance polynomial approximation and rank-reduced SVD algorithm can be effectively compared to the performances of other SVD-based methods in uniform sampling case. We will consider here two such methods which can be employed for parameter estimation when noise-corrupted signal samples $\{x[k] = x(kT); k=0,\cdots,L-1\}$ are provided at uniform spacing.

*Autocorrelation-like Matrix Method*

The autocorrelation-like matrix (ALM) method is an extension of the SVD-Prony method in the second-order statistics domain [7], [8]. The developed method provides estimation of signal parameters with less sensitivity to the effect of noise. A brief description of the ALM matrix method is given below.

The linear prediction equations satisfied for the signal given by (1) are expressed in matrix form as,

$$\mathbf{X} A = X \quad (10)$$

where $\mathbf{X}$ and $X$ are defined similar to $\mathbf{Y}$ and $Y$ respectively, as in (7) and (9). Then, instead of utilizing the rank-$M$ pseudoinverse of $\mathbf{X}$ to compute $A$, first both sides of (10) are processed to obtain,

$$\mathbf{R} A = R \quad (11)$$

where $\mathbf{R} = \left[ (1/L) \sum_{l=0}^{L-J-i-1} x[J+l+i-j] x^*[l]; i=1,\cdots,I; j=1,\cdots,J \right]$,

and $R^T = \left[ (1/L) \sum_{l=0}^{L-J-i-1} x[J+l+i] x^*[l]; i=1,\cdots,I \right]$,

and the actual value of $I$ should be considerably less than its maximum value $I_{max} = L - J - 1$. The rank-$M$ pseudoinverse of $\mathbf{R}$ is employed now in determining $A$, and once the coefficients are known, the characteristic equation is solved for its roots.

The noise desensitization of the developed method is achieved because first the matrix $[\mathbf{R}|R]$ can be separated into the autocorrelation-like signal matrix and autocorrelation noise matrix. Furthermore, the autocorrelation noise matrix tends to be the null matrix as the data length L approaches infinity. We assume here that the additive noise is white.

*Matrix Pencil Method*

The matrix pencil (MP) method has its root in the pencil-of function method which has been extensively used in system identification [9], [10]. The improved method while extracting the signal pole information from a generalized singular-value problem, utilizes extended order modeling. A truncated-SVD algorithm reduces the dimension of the final singular-value problem back to the true model order. As a result, the noise sensitivity of estimation is substantially reduced, and since only the signal poles are estimated, separation of signal and noise poles becomes unnecessary.

In the MP method, two matrices are formed, $\mathbf{X}$ as in (10), and $\mathbf{X}_1$ given by $\mathbf{X}_1 = \left[ x[J+i-j]; i=1,\cdots,L-J; j=1,\cdots,J \right]$. The generalized singular-value equation is then written as

$$\mathbf{X}_1 \mathbf{q} = z \mathbf{X} \mathbf{q} \quad (12)$$

which is solved for the generalized singular values. In the no-noise case, the $M$ non-zero generalized singular-values can be shown to be the $M$ $z_i$-poles of the signal.

The generalized singular-value problem can be reduced to a singular-value problem by premultiplying both sides of (12) with $\mathbf{X}^+$ which is the rank-$M$ pseudoinverse of $\mathbf{X}$,

$$\mathbf{X}^+ \mathbf{X}_1 \mathbf{q} = z \mathbf{X}^+ \mathbf{X} \mathbf{q} = z \mathbf{q} \quad (13)$$

Therefore, the singular-values of the matrix $\mathbf{X}^+\mathbf{X}_1$ are same as the $M$ signal and $(J-M)$ noise $z_i$-poles. The rank-$M$ pseudoinverse of $\mathbf{X}$ is defined as,

$$\mathbf{X}^+ = \sum_{i=1}^{M} (1/\sigma_i) \mathbf{v}_i \mathbf{u}_i^H = \mathbf{V} \mathbf{\Sigma}_M^{-1} \mathbf{U}^H \quad (14)$$

where $\{\sigma_i; i=1,\cdots,M\}$ are the $M$ largest singular values, $\mathbf{v}_i$'s and $\mathbf{u}_i$'s are the corresponding singular vectors, $\mathbf{V} = [\mathbf{v}_1,\cdots,\mathbf{v}_M]$, $\mathbf{U} = [\mathbf{u}_1,\cdots,\mathbf{u}_M]$, and $\mathbf{\Sigma}_M = \text{diag}[\sigma_1,\cdots,\sigma_M]$.

Substituting (14) into (13) and simplifying, it can be shown that the estimates of signal $z_i$-poles can be found by computing the eigenvalues of the $M \times M$ nonsymmetrical matrix, $\left( \mathbf{\Sigma}_M^{-1} \mathbf{U}^H \mathbf{X}_1 \mathbf{V} \right)$.

**Comparison of Three SVD-based Methods**

The comparison is valid only in the uniformly sampling case. Since in this case, the question of reconstructing the signal values at uniform spacing does not arise, the purpose of processing of data by minimum-variance polynomial approximation is to be explained.

It is not difficult to show that the autocorrelation-like matrix method will have the asymptotic properties of optimal estimation, provided the embedded noise is known to be white and Gaussian. For the method based on minimum-variance polynomial approximation, no such assumptions are needed. The fact that the preprocessing of data will make the remnant error closely Gaussian is the most desirable feature of the proposed method, which makes the designed estimator perform like an optimal estimator even in a realistic situation where the noise process is neither Gaussian nor white. This explains why preprocessing of data by the polynomial approximation is suggested even in the case where the signal samples are given at uniform spacing.

Among other advantages of the proposed method are significant enhancement of SNR level in the processed data, which reduces the effect of ill-conditioning of model equation in estimation [1], [4], and capability of handling the problem when the original noise process is known to be correlated and colored.

Compared to the first two SVD-based methods, it is observed that the MP method essentially leads to a suboptimal estimator. The main advantage of the MP method is the reduced dimension in the final eigenvalue problem, which leads to less computation and as a result, less computational error. The method possesses high efficiency in implementation because only the signal poles are computed here.

**Simulation Results**

*Example 1: Nonuniform sampling case*

The real-valued signal of desired form,

$$g(t) = \exp(s_1 t) + \exp(s_1^* t) + \exp(s_2 t) + \exp(s_2^* t),$$

where $s_1 = -1/75 + j2\pi 0.08$ and $s_2 = -1/90 + j2\pi 0.11$, corrupted with zero-mean white Gaussian noise is sampled at 50 nonuniformly spaced points which are chosen arbitrarily except that no sampling interval exceeds 1.1 unit. By applying minimum-variance polynomial approximation, the reconstructed signal values $y(nT)$ are computed at uniform interval of $T = 0.5$, as shown in Figure 1. The polynomial order is chosen to be 19 from the error-variance, $\sigma_N^2$ plot of Figure 2. Observe how the correlated error process $e(t)$ differs from the uncorrelated noise process $w(t)$, both plotted in Figure 3.

By setting the SNR level at 10, 20 and 40 dB, the magnitude of bias and the variance of estimates of each parameter are numerically computed from 100 independent realizations of noise sequences. The results are shown in Table 1. The extended model order is chosen to be 16 for optimal performance in all the cases.

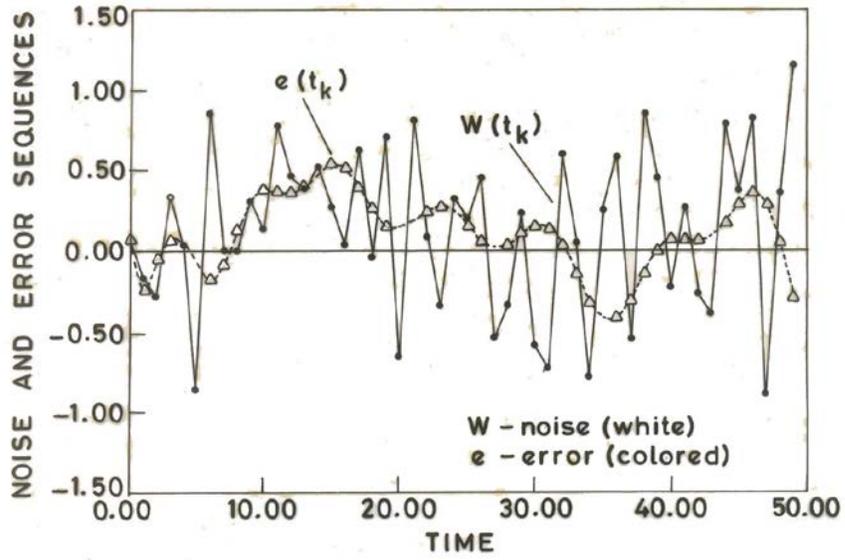

Figure 1 Orthogonal polynomial approximation: Peak SNR = 5 dB

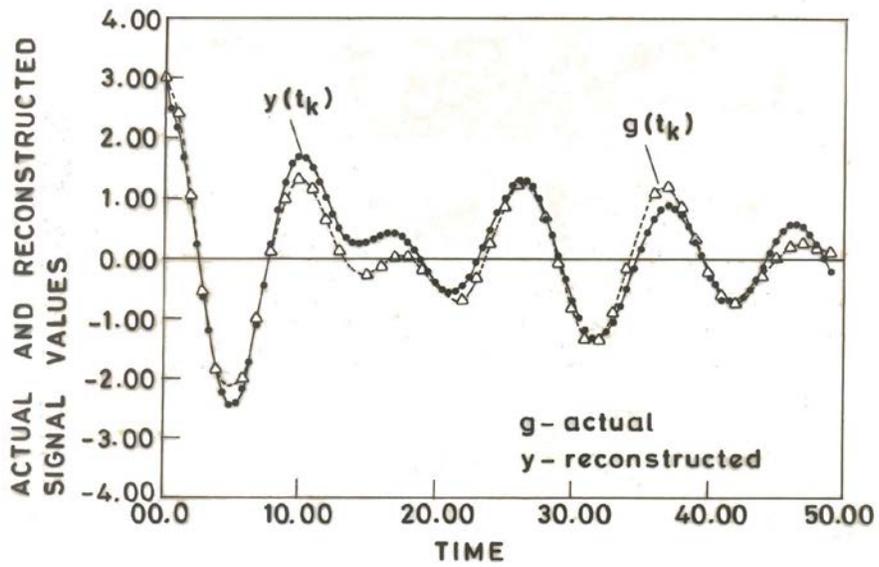

Figure 2 Minimum error-variance criterion: Approximation order = 19

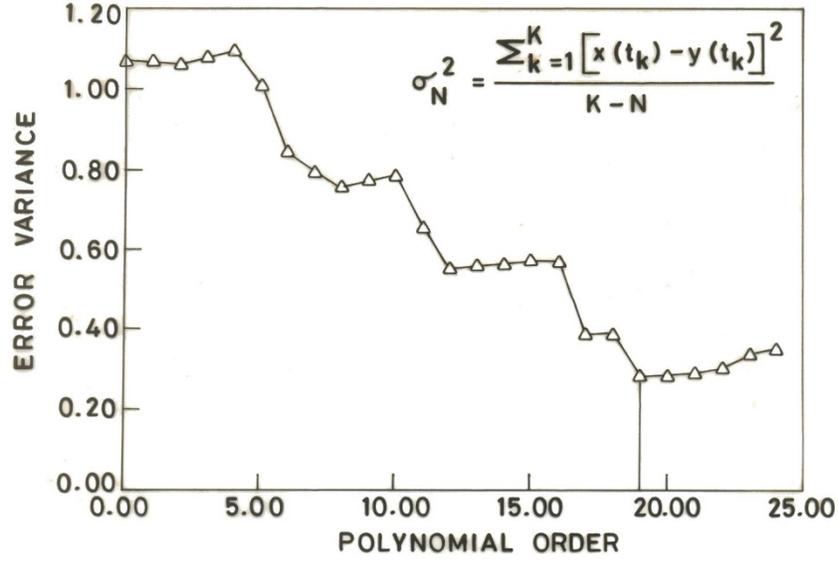

Figure 3 Preprocessing by minimum-variance polynomial approximation

Table 1 Bias and Variance of Parameter Estimates

| Estimated Parameters | SNR = 40 dB | | SNR = 20 dB | | SNR = 10 dB | |
|---|---|---|---|---|---|---|
| | Bias | Variance | Bias | Variance | Bias | Variance |
| $\alpha_1$ | $2.576 \times 10^{-4}$ | $1.174 \times 10^{-7}$ | $7.880 \times 10^{-4}$ | $8.303 \times 10^{-6}$ | $1.418 \times 10^{-3}$ | $1.409 \times 10^{-5}$ |
| $f_1$ | $7.659 \times 10^{-6}$ | $2.675 \times 10^{-8}$ | $8.240 \times 10^{-5}$ | $2.360 \times 10^{-7}$ | $1.431 \times 10^{-4}$ | $2.353 \times 10^{-6}$ |
| $\alpha_2$ | $4.118 \times 10^{-5}$ | $2.859 \times 10^{-7}$ | $6.330 \times 10^{-5}$ | $2.235 \times 10^{-6}$ | $5.521 \times 10^{-4}$ | $7.067 \times 10^{-6}$ |
| $f_2$ | $5.677 \times 10^{-5}$ | $5.772 \times 10^{-9}$ | $1.492 \times 10^{-4}$ | $2.649 \times 10^{-7}$ | $2.621 \times 10^{-4}$ | $1.027 \times 10^{-6}$ |

*Example 2: Uniform sampling case*

In this example, the transient signal

$$g(t) = \beta_1 \exp(s_1 t) + \beta_1 \exp(s_1^* t) + \beta_2 \exp(s_2 t) + \beta_2 \exp(s_2^* t),$$

with $s_1 = -0.00555 + j0.08$, $\beta_1 = 1.5$, $s_2 = -0.00666 + j0.11$, $\beta_2 = 3.5$, is sampled at 50 equispaced points with interval 5.6 units. The SNR value is set at 5 dB by mixing zero-mean white Gaussian noise sequence generated by computer.

The magnitude of bias and the variance of estimation for each parameter are computed from 200 independent realizations, by applying in turn the proposed method, the ALM method and the MP method. The comparative results, as shown in Table 2, agree well with the discussions presented in Section 3 where it is argued that the proposed method will be superior in accuracy of estimation.

Table 2 Comparison of Three SVD-based Methods

| Estimated Parameters | Our Method | | ALM Method | | MP Method | |
|---|---|---|---|---|---|---|
| | Bias | Variance | Bias | Variance | Bias | Variance |
| $\alpha_1$ | $1.006 \times 10^{-4}$ | $9.298 \times 10^{-5}$ | $4.704 \times 10^{-3}$ | $7.333 \times 10^{-4}$ | $6.583 \times 10^{-3}$ | $9.973 \times 10^{-3}$ |
| $f_1$ | $8.469 \times 10^{-4}$ | $7.983 \times 10^{-5}$ | $3.582 \times 10^{-3}$ | $1.301 \times 10^{-4}$ | $8.790 \times 10^{-3}$ | $8.049 \times 10^{-3}$ |
| $\alpha_2$ | $3.993 \times 10^{-4}$ | $1.820 \times 10^{-5}$ | $6.589 \times 10^{-3}$ | $2.518 \times 10^{-5}$ | $6.934 \times 10^{-3}$ | $3.346 \times 10^{-4}$ |
| $f_2$ | $3.212 \times 10^{-4}$ | $1.298 \times 10^{-5}$ | $4.114 \times 10^{-3}$ | $2.507 \times 10^{-5}$ | $7.063 \times 10^{-3}$ | $2.889 \times 10^{-4}$ |

For the MP method, the extended model order is set at 16 to provide best results. For the other two methods, the order is optimally chosen to be 20.

**Concluding Remarks**

We have proposed a complete approach based on minimum-variance polynomial approximation, maximum likelihood estimation and SVD algorithm. The proposed method provides very accurate estimates of parameters of complex exponentials under general conditions, viz., when the signal samples are not necessarily at uniform spacing, and/or the superimposed noise process may not be white or Gaussian.

The superiority of the proposed method is demonstrated by comparing its performance with the performance of the SVD-based autocorrelation-like matrix method and matrix pencil method in the uniform sampling case. Better results are obtained by the proposed method even when the noise process is white and Gaussian. Therefore, we may conclude that the proposed method with pre-processing of data should be the choice for better accuracy in all cases, although it needs some extra computation.

**References**

[1] Sircar, P.: Accurate Parameter Estimation of Damped Sinusoidal Signals Sampled at Nonuniform Spacings. PhD Dissertation, Elec. Eng., Syracuse University (1987)

[2] Sircar, P., Sarkar, T.K.: System Identification from Nonuniformly Spaced Signal Measurements. Signal Processing. 14(3), 253–268 (1988)

[3] Ranade, A.C.: System Identification from Nonuniformly Sampled Data. MTech Thesis, Elec. Eng. Dept., IIT Kanpur (1990)

[4] Sircar, P., Ranade, A.C.: Nonuniform Sampling and Study of Transient Systems Response. IEE Proc. F. 139(1), 49–55 (1992)

[5] Kay, S.M.: Modern Spectral Estimation: Theory and Application. Prentice Hall, Englewood Cliffs, N.J. (1988)

[6] Deprettere, E.F. (ed.): SVD and Signal Processing: Algorithms, Applications and Architectures. Elsevier Science, North-Holland (1988)

[7] Kumaresan, R., Tufts, D.W.: Estimating the Parameters of Exponentially Damped Sinusoids and Pole-Zero Modeling in Noise. IEEE Trans. on Acoust., Speech, Signal Processing. 30(6), 833–840 (1982)

[8] Cadzow, J.A., Wu, M.M.: Analysis of Transient Data in Noise. IEE Proc. F. 134(1), 69–78 (1987)

[9] Jain, V.K., Sarkar, T.K., Weiner, D.D.: Rational Modeling by Pencil-of-Functions Methods. IEEE Trans. on Acoust., Speech, Signal Processing. 31(3), 564–573 (1983)

[10] Hua, Y., Sarkar, T.K.: Matrix Pencil Method for Estimating Parameters of Exponentially Damped/Undamped Sinusoids in Noise. IEEE Trans. on Acoust., Speech, Signal Processing. 38(5), 814–824 (1990)

**Further reading**

- Ravi Shankar, M., Sircar, P.: Nonuniform sampling and polynomial transformation method. In Proc. IEEE International Conference on Communications, ICC 2002, 3, 1721–1725 (2002)

- Banoth, J.K., Sircar, P.: Polynomial transformation method for non-Gaussian noise environment. *arXiv preprint arXiv:1401.5580* (2014)

- Sircar, P.: Mathematical Aspects of Signal Processing. Cambridge University Press (2016)